\documentclass[aps,showpacs,twocolumn,floats,floats,superscriptaddress,floatfix, longbibliography]{revtex4-1}
\usepackage{amssymb}
\usepackage{graphicx}
\usepackage{bm}
\usepackage{amsfonts}
\usepackage{color}
\usepackage{amsmath}    
\usepackage{epsfig}
\usepackage{hyperref}   
\usepackage{mathrsfs}

\newcommand{\ictsaddress}{International Centre for Theoretical Sciences, Tata Institute of Fundamental Research, Bangalore 560089, India}
\newcommand{\iitaddress}{Department of Chemical Engineering, Indian Institute of Technology Bombay,
Mumbai 400076, India}
\newcommand{\nordita}{Nordita, KTH Royal Institute of Technology and Stockholm University, Roslagstullsbacken 23, 10691 Stockholm, Sweden}
\begin{document}
\title{Fluid dynamics in clouds: The sum of its parts}
\author{S. Ravichandran}
\email{ravichandran@su.se}
\affiliation{\nordita}
\author{Jason R. Picardo}
\email{jrpicardo@che.iitb.ac.in}
\affiliation{\iitaddress}
\author{Samriddhi Sankar Ray}
\email{samriddhisankarray@gmail.com}
\affiliation{\ictsaddress}
\author{Rama Govindarajan}
\email{rama@icts.res.in}
\affiliation{\ictsaddress}

\begin{abstract}
This paper is aimed at describing cloud physics with an emphasis on fluid dynamics. As is inevitable for a review of an enormously complicated problem, it is highly selective and reflects 
of the authors' focus. The range of scales involved, and the relevant physics at each scale is described. 
Particular attention is given to droplet dynamics and growth, and turbulence with and without thermodynamics.
\end{abstract}
\maketitle

\section{Glossary }

Aerosol: tiny ($\sim 0.1-1$ micron) solid particles suspended in the air.
There are about 100-1000 aerosol particles per cubic centimeter of air.

Clouds: mixtures of air ($\sim 99\%$ by weight), water vapour (1\%), liquid
water droplets (0.1\%), aerosol particles, trace gases. Clouds are usually in turbulent flow.

Caustics: regions of the flow where particles with different velocities
arrive simultaneously at the same location.

Supersaturation: a system that has more water vapour than the saturation value prescribed
by the Clausius-Clapeyron equation \ref{eq:clausius}.

Ventilation effects: the effects of oncoming flow on the growth of
droplets. Used in the context of water droplets growing by condensation.

\section{Why study clouds}

Clouds, since they involve many different phenomena interacting with each other
in complex ways, are of interest purely from scientific curiosity.
For instance, is it possible to predict what cloud shapes will result for given
atmospheric conditions? More importantly, perhaps, clouds are also immensely
influential in the energy and mass balances in the planet's atmosphere. In fact,
clouds are the last great sources of uncertainty in climate science.

Clouds increase the planet's albedo, reflecting away sunlight before
it can make it to the surface; they also act to provide a greenhouse
effect, trapping energy radiated away from the surface. These two opposing
effects are both of much larger magnitudes than any other sources in the
radiative balance of the planet (\cite{archer2011} chapter 6, \cite{stevens2013}.
The response of clouds to a warming planet---whether clouds will act
to slow down or to accelerate the planet's warming---is not clear at 
present (although very recent studies (\cite{schneider2019}) suggest that they
do indeed act as positive feedback). This uncertainty is due to the large magnitudes
of the aforementioned effects, and the fact that clouds are coupled with the global
circulation. 

The selective annual Northward propagation of the cloud-band known as the ITCZ (Inter-Tropical Convergence Zone) over longitudes including those of the Indian landmass, brings the Indian monsoon, among the biggest weather events, which provides fresh water for close to two billion people. Understanding the dynamics of the ITCZ requires understanding the dynamics of clouds, which is as yet an open problem (\cite{bony2015}). These are only a few of the most compelling reasons to study clouds.

With the advent of machine learning and associated statistical and
data-driven techniques, and the increasing availability of dedicated
computing power, it is tempting to rely solely on such statistical
methods. However, understanding the dynamics is useful not just as
a scientific exercise but also pragmatically.
Statistical techniques---machine learning in particular---are best
used in scenarios for which they have been `trained'. Most estimates 
suggest that the feedback from clouds on the climate is likely to
affect the circulation in the atmosphere substantially. The resulting 
large changes in the dynamics may not be possible to capture with 
machine learning techniques. The planet needs us to study the
dynamics of clouds! (\cite{schiermeier2015}.) \\

\section{Definition of the Subject }

The fluid dynamics in clouds covers length scales from tenths of microns
to hundreds of kilometres. Being a nonlinear problem, the physics at each scale has an effect on other scales. There are open questions which require an understanding of the basic physics at each scale, and also in the connections between scales. The lower end of this range concerns the
chemistry and chemical physics of aerosols. Aerosols are crucial to cloud formation, because they act as nuclei for droplet formation, as will be discussed below. Aerosols are introduced
into the atmosphere in a variety of ways, natural and anthropogenic.
The production of aerosols, especially sea-salt aerosol by the mechanics of wave-breaking
at the surface of the ocean, is an outstanding problem of fluid mechanics.
The upper end of the range of length scales covers the dynamics of
weather and the climate. Significant progress has been made in recent
years, aided by advances in supercomputing, in the ability to make
reasonable predictions of the dynamics on these scales. The robustness
of these predictions depends, however, on understanding the global dynamics
at the `sub-grid scales'. The intermediate range, incorporating the
interactions of buoyancy-driven fluid turbulence of (dilute) suspensions,
is not only exceedingly complex, but also controls the dynamics of
processes at the largest scales related to weather and climate. We
will describe recent progress in understanding the dynamics in the
intermediate length scales.

Studies of the cloud dynamics can be based on observations of
clouds either in the real world or in the laboratory, or
on analyses or numerical solutions of the fluid dynamical equations
of motion. Our work is in the latter, and we will for the most
part restrict ourselves to discussing theoretical/ numerical studies
of clouds, although we do make note of some relevant experimental
studies. 

An important ingredient in the intermediate scales is that clouds are usually in turbulent flow. Turbulence consists of vortices and regions of shear whose length scales span a large range, starting from the biggest scale in a single cloud, of the order of a few kilometres, to what is known as the Kolmogorov scale $\eta$, which is of the order of a millimetre in a cloud. A turbulent flow is characterised, among other properties, by its Reynolds number, which is a ratio of inertial and viscous forces.
Consider a cloud of length scale $L \sim 1$ kilometre in height and width, where velocities $U$ are of the order of $10$m/sec. The kinematic viscosity $\nu$ of air is about $10^{-5}m^2/$sec. The Reynolds number is $LU/\nu \sim 10^9$.

The range of length scales involved even within the intermediate range in clouds is vast. Accurate
direct numerical simulations (DNSs), solving the Navier-Stokes equations or their variants, have to resolve the Kolmogorov scales of turbulence. If these scales have to be resolved in simulations
of a cloud of the length scale of 100m, each dimension has to be resolved
with $O(10^{5})$ grid points. Such numerical simulations are impossible
with today's computing resources. DNSs of cloud flows are typically conducted only within small boxes,
 of a few metres in length   \cite{kumarshaw2012, kumarshaw2013,kumarshaw2014}. In other words a small volume within a cloud is all we can simulate. In effect experiments (on a computer or in the laboratory) can be performed at Reynolds numbers that are much smaller than those found in clouds, in the hope that this will nevertheless provide useful answers \cite{abma2013,delozar2013,delozar2015,delozar2017, pauluis2010, schumacher2010moistrbc, weidauer2010, pauluis2011, RN2011PNAS,ravichandran_narasimha2019} at cloud Reynolds numbers (see also section \ref{subsec:Cumulus}). Workarounds for this limitation take the form of large eddy simulations (LESs) which resolve only the large scales of motion (i.e. they are `cloud-resolving') \cite{randall2003, jarecka2009, romps2010entrainment, grabowski2014} and use models to account for the smaller scales including the microphysics of phase-change.

The radius $a$ of a typical water droplet in a cloud ranges from about a micron to a few millimetres. Obviously even simulations that resolve the Kolmogorov scales of the
flow cannot resolve the scales associated with the motion of the water
droplets in clouds. Even the simplest approach to tracking droplets adds significantly to the burden of computations, given that there are $O(1000)$ small droplets per cubic centimetre of cloud. In the simplest approach, the finite-sized water droplets have to be treated as point particles
and tracked in a Lagrangian sense. Alternatively, these particles
can be coarse-grained into a field. The relative efficacies of these
two approaches to particle-dynamics are studied in \cite{prasad_dhruba2018}. The effects of finite droplet size and how this changes
their dynamics is discussed at length in section \ref{subsec:particles_caustics_collisions}. These are known as ``one-way coupled'' approaches, where the fluid equations are solved for without taking into account the fact that fluid is carrying particles and droplets, whereas the dynamics of the particles and droplets are dictated by the fluid motion. For a dilute suspension of small particles and droplets this is a fair approach. However, larger raindrops can affect the flow and can affect the dynamics of each other, and a perfect treatment would have to account for the forces of these objects on the fluid and on each other (two-way or four-way coupling). This can make computational costs forbidding. 

The thermodynamics taking place within a cloud has an important effect on the dynamics. Phase change results in heat release, which results in buoyancy. The potential energy thus gained is converted into the kinetic energy of turbulence. Thus turbulence in a cloud is fundamentally different from mechanically forced turbulence, and turbulence from heat supplied at the boundaries, which are studied most often. We return to this point in section \ref{subsec:Thermodynamics-of-phase-change}.  The droplet-growth bottleneck is a well-known open problem. Droplets can grow quickly to about 10 microns in size in a supersaturated environment typical of clouds. Once they are about $50$ microns in diameter, gravity can aid in the process of droplet growth by enhancing collision probability, with some fraction of all collisions resulting in coalescence. How droplets grow from about $10$ to about $50$ microns is not completely understood yet, and this is known as the droplet-growth bottleneck. Turbulence is widely accepted now to be a big part of the answer, and this is discussed below.

Most present-day studies assume that water droplets are spherical in shape whereas larger drops are sensitive to gravity and can distort in shape from the spherical, even to the point where they adopt shapes which are locally sheet-like, which then causes breakup into smaller droplets. Ice crystals are most often not spherical. Thus, the roles of shape, surface tension, gravity and even surface chemistry on the dynamics have to be studied. These effects are also important in
the thermodynamics of water droplet growth (discussed in section \ref{subsec:Thermodynamics-of-phase-change}) and in the collisions-coalescence of water droplets (section \ref{subsec:particles_caustics_collisions}).

Another important attribute is that over the length-scales that clouds occupy in the atmosphere, air is a compressible fluid. Note that the Earth's atmosphere at a height of ten kilometres is only a tenth as dense as it is on the ground. So a parcel of air, as it rises, will undergo significant expansion, which cannot be neglected by assuming incompressibility. The equations of compressible fluid motion are significantly
more complicated than those of an incompressible fluid (which itself
is a ``Millennium problem"). The fully compressible equations of motion for air contain sound waves which operate on very short timescales. These sound waves are unimportant for the dynamics of interest to usand computing them would require short timesteps and greatly increase the computational requirements. Fortunately, since the Mach numbers $Ma$ associated with the flow are small, the thermodynamic pressure of the ambient can be decoupled from the pressure fluctuations due to the motion (which are $O(Ma^2)$ relative to the thermodynamic means). This allows the use of the anelastic equations for flows over large heights or the incompressible equations for shallow flows, the latter of which is the limit we are concerned with. 
A sketch of the derivation of the anelastic and incompressible equations from the
compressible equations may be found in \cite{durran1989improving,bannon1996anelastic}.
As the name suggests, the anelastic equations `filter out' the sound waves
from the dynamics, leaving only the effects of compressibility on
the large scale dynamics. On relatively small scales, the assumption
of incompressibility is reasonable and is typically made in studies
of the flows in shallow clouds \cite{pauluis2010,schumacher2010moistrbc, weidauer2010, pauluis2011},
and even in some idealised studies of deep convection \cite{hernandez2013minimal}. \\

\begin{figure}
\includegraphics[width=0.75\columnwidth]{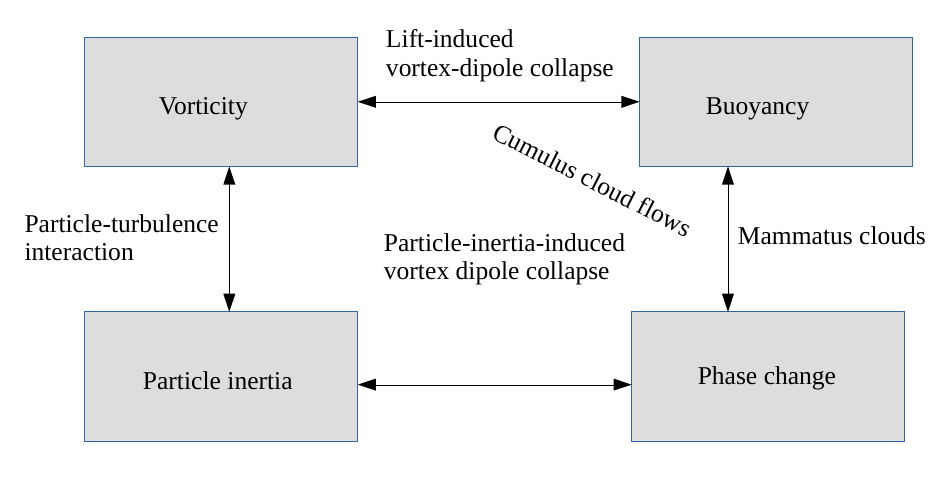}
\caption{\label{fig:interactions}Cloud dynamics as the sum of its components.}
\end{figure}
As we see in figure \ref{fig:interactions}, the dynamics of clouds
is an interplay of particle inertia, thermodynamics, the resulting
buoyancy driven flow. At the largest scales, the effect of Earth's rotation, and solar radiation and its modification by clouds, need to be understood better, and we do not deal with these topics in the present paper.

In summary, studies at different scales have to sometimes be carried out in isolation, using approaches and assumptions appropriate for that scale. New physics is revealed at each scale, and their effects must then be included in our studies at other scales.

\section{Microphysics without thermodynamics \label{subsec:particles_caustics_collisions}}

\begin{figure*}
\includegraphics[width=1.0\textwidth]{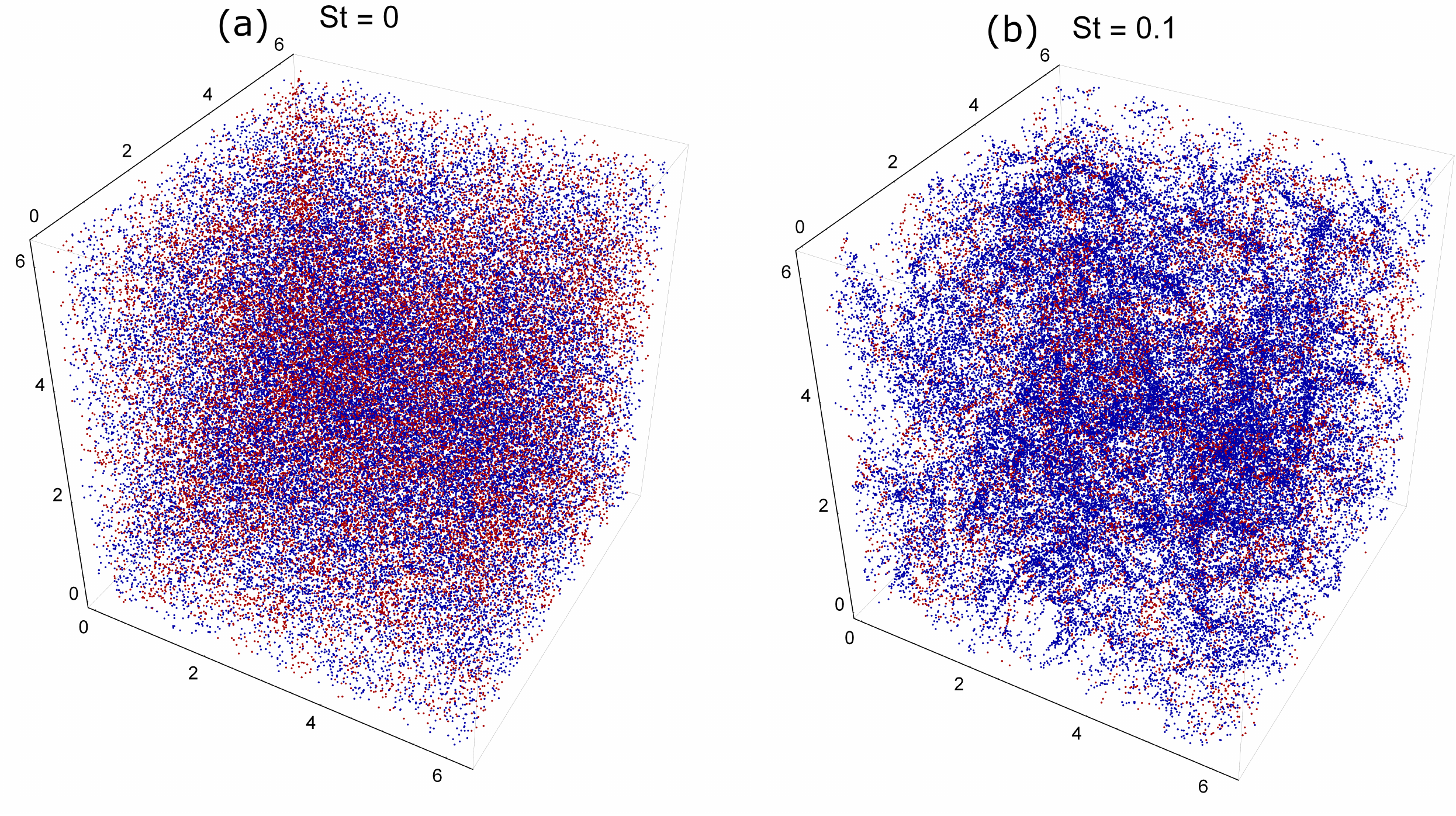}
\caption{Snaphots of tracers (a) and inertial particles (b), with $St = 0.1$, in a three-dimensional turbulent flow. Particles in regions dominated by rotation ($\mathcal{Q}>0$) are colored red, while those in regions dominated by straining ($\mathcal{Q}<0$) are colored blue. The dissipative dynamics of inertial particles ($St = 0.1$) causes them to form dynamic clusters, which are seen in panel (b) to mainly reside in straining regions, in accordance with the ejection of inertial particles from rotational zones.}
\label{3Dcluster}
\end{figure*}

A simple framework, which ignores the  effects of thermodynamics, phase changes
and the associated changes in temperature, to understand the physics of a
\textit{single} warm cloud is to model it as a dilute suspension of small,
spherical water droplets of radius $a$ which are advected by a statistically
stationary, homogeneous and isotropic, full-developed turbulent flow. Such an
approach ignores the effect of condensation, arising from a super-saturated
environment, by assuming that the starting point of such studies are non-precipitating
droplets which are already condensed to sizes of about 10$\mu m$; hence further
growth through condensation over a reasonably short time window, corresponding
to the life-time of such a cloud, is unlikely~\cite{Shaw-Review-2003,Grabowski-review-2009,Lanotte-JAS-2009,Bodenschatz-Editorial-2010,Devenish-review-2012}. 

Such a simplification has at least two distinct advantages. Firstly, it allows
us to formulate and address questions of collisions, coalescences, and
gravitational settling (precipitation) in the turbulent setting of a cloud in a
precise way. Secondly, given this framework, it lends itself easily to the use
of tools and ideas developed in the field of turbulent transport over the last
two decades or so. 

In typical clouds, given $a/\eta \ll 1$, the Reynolds number associated with a
droplet $Re_p \ll 1$. This allows us to define the dynamics of a droplet, in
the presence of a gravitational force ${\bf g}$ and an (turbulent) advecting
fluid velocity ${\bf u}$, in terms of its position ${\bf x}_p$ and velocity ${\bf
v}$, through the linearised Stokes drag model with a Stokes time $\tau_p$~\cite{CroorRev-2017}:
\begin{subequations}\label{eq:DNS}
\begin{align}
\label{eq:sta}	\frac{d{\bf x}_p}{dt} &= {\bf v}; \\
\label{eq:stb}	\frac{d{\bf v}}{dt} &= -\frac{{\bf v} - {\bf u}({\bf x}_p)}{\tau_p} + {\bf g}.
\end{align}
The velocity field of the carrier flow, driven to a statistically steady state through a force ${\bf f}$, 
with density $\rho_f$, a kinematic viscosity $\nu$, and a pressure field $P$, 
satisfies the incompressible, three-dimensional Navier-Stokes equation
\begin{align}
\label{eq:NSa}	\frac{\partial {\bf u}}{\partial t} + ({\bf u}\cdot \nabla)&= \nu \nabla^2{\bf u} - \frac{\nabla P}{\rho_f} + {\bf f}; \\
\label{eq:NSb}	\nabla \cdot {\bf u} &= 0. 
\end{align}
\end{subequations}
Given the assumptions of a small droplet and a dilute suspension, the underlying flow is assumed to be unaffected 
by the presence of such water droplets.

The effect of the finite size and the density contrast of the particle with the
carrier flow, which leads to a finite time of relaxation of the particle
velocities to that of the fluid, is captured by the Stokes time $\tau_p =
\frac{2a^2\rho_p}{9\nu \rho_f}$, where the particle density is given by
$\rho_p$; for clouds (water droplets in air), the ratio of these two densities 
is $\rho_p/\rho_f \sim 1000$. However, it is useful to measure this inertia of
the particles in terms of the non-dimensional Stokes number $St =
\tau_p/\tau_\eta$, where $\tau_\eta = \sqrt{\nu/\epsilon}$ is the
characteristic, short-time, Kolmogorov time-scale of the fluid ($\epsilon$ is
mean energy dissipation rate). Such non-dimensional numbers allow an easy comparison 
between observations, experiments, theory and numerical simulations.

The linear Stokes drag model (Eqs.~\ref{eq:sta}-\ref{eq:stb}) is, of course, in the heavy-particle limit $\rho_p \gg \rho_f$,
a simplification of the Maxey-Riley equation~\cite{Maxey-Riley-1983} for the motion of a spherical particle (with $Re_p \ll 1$) in a flow:
\begin{widetext}
\begin{equation}
\rho_{p} \frac{d{\bf v}}{dt}  =  \rho_{f} \frac{D{\bf u}}{Dt} + (\rho_{p}-\rho_{f}) {\bf g}
- \frac{9 \nu \rho_{f}}{2a^2}\left({\bf v}-{\bf u}-\frac{a^2}{6} \nabla^{2} {\bf u}\right)
- \frac{\rho_{f}}{2}\left(\frac{d {\bf v}}{dt} - \frac{D}{Dt}\left[{\bf u} + \frac{a^{2}}{10} \nabla^{2} {\bf u}\right]\right)
- \frac{9 \rho_{f}}{2a} \sqrt{\frac{\nu}{\pi}} \int^t_0  \frac{1}{\sqrt{t-\xi}}\frac{d}{d\xi}({\bf v}-{\bf u}-\frac{a^{2}}{6}\nabla^{2}{\bf u}) \mathrm{d}\xi 
\label{maxey}
\end{equation}
\end{widetext}
where $\frac{D}{Dt}$ denotes the full convective derivative and it factors in the effects of the force due to the 
undisturbed flow $\rho_{f} \frac{D{\bf u}}{Dt}$, the buoyancy $(\rho_{p}-\rho_{f}) {\bf g}$, the 
Stokes drag $\frac{9 \nu \rho_{f}}{2a^2}\left({\bf v}-{\bf u}-\frac{a^2}{6} \nabla^{2} {\bf u}\right)$, 
the added 
mass $\frac{\rho_{f}}{2}\left(\frac{d {\bf v}}{dt} - \frac{D}{Dt}\left[{\bf u} + \frac{a^{2}}{10} \nabla^{2} {\bf u}\right]\right)$, 
and the Basset history $\frac{9 \rho_{f}}{2a} \sqrt{\frac{\nu}{\pi}} \int^t_0  \frac{1}{\sqrt{t-\xi}}\frac{d}{d\xi}({\bf v}-{\bf u}-\frac{a^{2}}{6}\nabla^{2}{\bf u}) \mathrm{d}\xi$ effects. Without going into a rigorous demonstration of how Eq.~\ref{maxey} reduces 
to Eq.~\ref{eq:stb}, we can immediately see that for heavy particles the force due to the undisturbed flow is negligible and the 
only effect of gravity is a net acceleration downwards. Furthermore, the Faxen corrections $\sim a^2\nabla^2{\bf u}$ for 
small particles are negligible as is the Basset term in such dilute suspensions of passive particles. (On the last aspect, we refer 
the reader to a recent work by Prasath \textit{et al.}~\cite{Ganga-JFM-2019} for a detailed analysis of the Basset history term). 
Indeed recent work by Saw \textit{et al.}~\cite{Saw-PoF-2014} have confirmed by comparing experimental data with those obtained 
from a numerical simulation of Eqs.~\ref{eq:DNS}, that the approximation discussed here are indeed reasonably valid for 
the dilute suspensions of small, but heavy, particles that we consider in this paper. 

Before we proceed further, it might be useful at this stage to comment on how
Eqs.~\ref{eq:DNS} are solved on the computer~\cite{Canuto-2006}. (See Ref.~\cite{Pandit-PoF-Review} for 
a discussion on the nature of such simulations in two-dimensional flows.) The incompressible
Navier-Stokes equations (Eqs.~\ref{eq:NSa}-\ref{eq:NSb}) are typically solved,
in three dimensions, on a triple-periodic $2\pi$ cube with $N^3$ collocation
points; in the results being reviewed in this paper, $N$ has ranged from 512 up
to 2048 yielding Taylor-scale based Reynolds number which range from
(approximately) 120 to 450. The flow is driven to a statistically steady,
homogeneous and isotropic, turbulent state with an external forcing.  There is
of course considerable freedom in the way we choose to force the fluid; two
particularly popular choices are one with a constant energy injection as low
wave-numbers~\cite{Sawford-1991} and another where (again at low-wavenumbers) a second-order
Ornstein–Uhlenbeck process~\cite{Pope-2005} is adapted to provide a more \textit{random}
forcing.  The equations themselves are solved through a standard
pseudo-spectral method where spatial derivates are taken in Fourier space to
allow an easy integration in time of what essentially becomes an algebraic
equation.

\begin{figure*}
\includegraphics[width=1.0\textwidth]{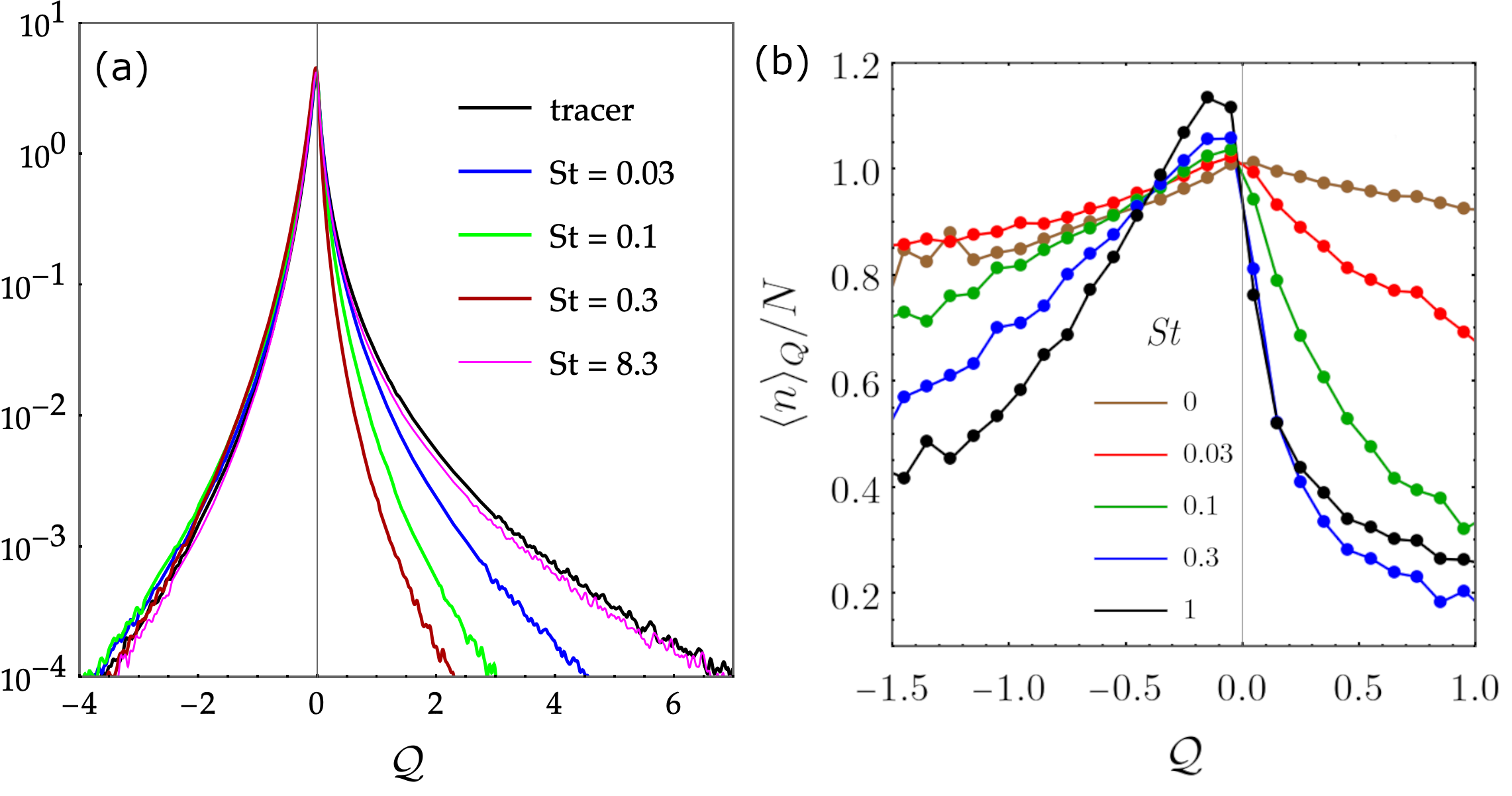}
\caption{Panel (a) presents PDFs of $\mathcal{Q}$ sampled by tracers and inertial particles with various values of $St$. Panel (b) shows the average coarse-grained particle number density, conditioned on the local value of $\mathcal{Q}$, for various $St$.}
\label{Qpdf_den}
\end{figure*}

Solving for the particle dynamics (Eqs.~\ref{eq:sta}-\ref{eq:stb}) are less
involved. The only non-trivial element in this is to use suitable interpolation
schemes to obtain the fluid velocity, which is calculated on a regular Eulerian
grid, at typically off-grid particle positions. Several such schemes exist and
the ones commonly used are the cubic spline, the B-spline, the trilinear, or the cubic
interpolation schemes (see e.g. \cite{vanHinsberg2013}). 

The consequences of the linear Stokes drag model have been studied
extensively~\cite{Bec-PoF-2005,Chun-2005,Bec-PRL-2007,Monchaux-2012,Gustavsson-2016,Ireland-2016}
since the  pioneering work of Bec~\cite{Bec-2003,Bec-JFM-2005}. A detailed
discussion of this is certainly beyond the scope of this present paper.
However, in what follows, it is useful to recall two central features of the
dynamics defined by Eqs.~\ref{eq:DNS}, namely \textit{preferential
concentration} and \textit{caustics}.

The dissipative dynamics of the particle motion leads to a preferential sampling 
of the flow and hence a preferential concentration of particles as opposed to 
a homogeneous distribution in the flow (as seen for tracers) for finite Stokes numbers: 
Particles with finite sizes evolve to (dynamically changing) attractors with fractal dimensions.
Traditionally, this preferential concentration or inhomogeneities in the distribution of 
particles is captured through the correlation dimension $D_2$ obtained from calculating 
the probability  $P^<(r)$ of two particles being within a distance $r$, whence 
$P^<(r) \sim r^{D_2}$. The correlation dimension $D_2$ is of course a function of the 
Stokes number $St$. In the limit of vanishing $St$ (tracers), particles must distribute 
homogeneously and hence, in a three-dimensional flow, $D_2 = 3$. For very large values of 
$St$, the motion of particles are essentially decorrelated from the flow and thus 
have a more ballistic behaviour. This results, yet again, in space-filling and consequently 
$D_2 = 3$. At intermediate values of $St$, however, the particles distributions are inhomogeneous $D_2 < 3$ and 
maximal clustering (or inhomogeneities) is observed (with an accompanying minimum in $D_2$) for 
$St \sim \mathcal{O}(1)$. We refer the reader to Figs. 1 and 2 in Ref.~\cite{Bec-PRL-2007} for 
an illustration of these effects.

From the perspective of flow structures, the clustering of heavy inertial
particles can be tied to their ejection from vortical or rotational regions of
the flow. One way to quantify this behaviour is via the
$\mathcal{Q}$-criterion~\citep{Dubief2000}, which uses the local velocity
gradient matrix $\mathcal{A}$ to define a quantity $\mathcal{Q} \equiv
(R_{ij}^2 - S_{ij}^2)/2$ (where $R = (\mathcal{A} + \mathcal{A}^T)/2$ and $S =
(\mathcal{A} - \mathcal{A}^T)/2$). Regions dominated by rotation (straining)
have $\mathcal{Q}>0$ ($<0$). Fig.~\ref{Qpdf_den}(a) presents the Lagrangian
probability distribution functions (PDFs) of $\mathcal{Q}$ measured along the
trajectories of tracers and inertial particles with various values of $St$. The
undersampling of vortical regions by inertial particles is clearly visible;
indeed the effect is significant even for $St$ as small as 0.03 (relevant for
$10-20$ $\mu$m cloud droplets). This effect gets stronger upto $St \approx
0.5$, beyond which the particles begin to decorrelate from the underlying flow
structures, until eventually, for large $St$, the PDF approaches that for
tracers (cf. $St = 8.3$ in Fig.~\ref{Qpdf_den}(a)).

The fact that ejection from vortices leads to concentration and clustering in
straining regions is shown clearly in Fig.~\ref{Qpdf_den}(b), which presents a
coarse-grained particle number density (using cubic bins of side 20$\eta$),
conditioned on the local value of $\mathcal{Q}$, and for various $St$. As $St$
increases from zero, the density of particles in regions of moderate straining
increases, while the density in rotational regions reduces. This is illustrated
visually by Fig.~\ref{3Dcluster}(b), wherein most of the dense particle
clusters are blue ($\mathcal{Q}<0$). Fig.~\ref{Qpdf_den}(b) also shows that
particle density reduces with $St$ in regions of very high straining,
indicating that inertial particles cannot cluster in such intense regions of
the flow, but rather prefer mildly straining zones.

A second important ingredient in this story is caustics
~\citep{Ravichandran2015}. Caustics are defined here as regions in the flow
where droplets of different velocities can arrive simultaneously at the same
location. In other words, caustics are regions of the flow where droplet
velocity cannot be described as a field. These are significant for the
following reason. Droplets of moderate Stokes number ($\sim 0.1$ to $1$) get
ejected quickly out of vortices, and this increases their propensity to collide
and coalesce, but smaller droplets are usually taken to behave like passive
scalars, just advecting passively with the flow. Under this assumption no
preferential concentration takes place for small droplets, and collisions would
be extremely rare. We thus do not have a mechanism by which small droplets can
coalesce and begin to cross the bottleneck. In the immediate vicinity of a
single vortex of circulation $\Gamma$, i.e., within a distance $\sim
\sqrt{\Gamma \tau_p}$ from the vortex centre, even the smallest droplets are
centrifuged out ~\citep{Ravichandran2015}, and caustics can form. It was shown
\citep{Ravichandran2015,Deepu2017} that collisions between small droplets can
be greatly enhanced in the caustics region, giving rise to a small number of
droplets which can cross the bottleneck, and become the seeds for further
coalescence events.

Given this context, let us return once more to questions within the framework
of turbulent transport which are pertinent to the problem of droplet dynamics
in a warm cloud. In particular, the questions that we discuss in this paper
have to do with collisions, coalescences, and precipitation. Specifically
these are best discussed by answering the following questions: How fast---and
where---do droplet collide? How fast do droplets grow (by coalescence)? And,
how fast do droplet settle under gravity? (The issue of the structure of such aggregates when 
they collide, but not coalesce, will not be covered in this overview~\cite{sticky,gupta}.)

Turbulence is thought to play a dominant role in enhancing the droplet-droplet
collision rates, and in turn the droplet-size distributions as well as the
initiation time of rain, in typical warm
clouds\cite{Shaw-Review-2003,Falkovich-2002}. The underlying
mechanism instrumental in this is not only preferential concentration,
discussed above, but also, through
\textit{slings}~\cite{Falkovich-2002,Bewley-2013} and
\textit{caustics}~\cite{Wilkinson-2006,Falkovich-2007}, the extreme
velocities with which particles can approach each other. Therefore in order to gain insights 
which can help build mesoscopic models for collision kernels, it is important to have reliable estimates 
of the typical relative velocities between droplets which are about to collide in a turbulent flow. 

This issue was addressed by Saw \textit{et al.}~\cite{Saw-PoF-2014} who, through experiments,
numerical simulations and theory, studied the probability distribution
functions of the velocity differences between pairs of particles, measured
along the line-of-sight, when they are quite close to each other.  In the
experiment, a turbulent flow was generated within a $1\;m$-diameter acrylic
sphere~\cite{Bewley-2013} with Taylor micro-scale Reynolds numbers $R_{\lambda}$
as high as 190 corresponding to values of $\eta$ as low as $180\;\mu m$. In
such a flow, a bi-disperse population of droplets were introduced, via a
spinning disc~\cite{Walton-49}, with mean diameters $6.8\mu m$ and 19$\mu m$
which are much smaller than $\eta$.  Given that the experiment was able to
achieve three different $R_{\lambda}$, and hence $\eta$, it was possible to
obtain particle trajectories for 6 different Stokes numbers through
stereoscopic Lagrangian Particle Tracking~\cite{Ouellette-2006}. 
In the numerical simulations, of the sort described above, particle trajectories were 
obtained for the similar Stokes numbers and for comparable values of $R_{\lambda}$ which 
allowed a meaningful comparison of theory and experiments.

A convenient measure of the statistics of how droplets impact on each other, is
through the probability distribution function of the longitudinal component of
the velocity differences $v^\parallel$ between pairs of particles and
conditioned on their separation $r$. In Fig. 1 of Ref.~\cite{Saw-PoF-2014}, these
distribution functions for four
different values of the Stokes number and, in each case, conditioned on three different 
values of the separation $r$ are shown. The agreement between the
experimental and numerical data, especially for the larger values of $St$, is a
confirmation of the validity of the linearised Stokes drag model of
Eqs.~\ref{eq:DNS}.  However, it is worth pointing out that these distributions,
which seem to fit the form of stretched
exponentials~\cite{Kailasnath-1992} and at odds with the compressed
exponential prediction for large Stokes numbers~\cite{Gustavsson-2008},
show consistently,  for reasons still not clear, a greater convergence between
the numerical and experimental data for the left tail (approaching pairs) than
for the right tail (separating pairs). Furthermore, a closer examination of
these distributions show that droplet-pairs approach each other with increasing
relative velocities---and a possible increase in collision rates---as their
Stokes number increases consistent with other evidences of the sling
effect~\cite{Bewley-2013}.

These distributions were of course conditioned on small ($\mathcal{O}(\eta)$)
but still finite separations; in the context of understanding collisions
amongst droplets in a cloud, we should examine the relative velocities at
contact or at least when they are even closer to each other ($r \to 0)$. It
turns out that all these distributions can be collapsed   on top of one another
(Fig. 2 in Ref.~\cite{Saw-PoF-2014}) by a simple rescaling with $r^\beta$;
experimental and numerical data suggests that $\beta$ has a mild
Stokes-dependence and for reasonably large values of $St$ (not entirely valid
for droplets in a cloud at its infancy), corresponding to extreme velocity
differences, its value is consistent to the saturation exponent $\xi_{\infty}$
of the higher-order moments of relative velocities~\cite{Bec-JFM-2010}.
The positive exponent $\beta$, which is small for small droplets, nevertheless 
suggests the possibility of mild impact velocities on contact.

The particle dynamics in a more realistic cloud is of course non-stationary as
droplets grow and change their sizes and numbers. A step in this direction is
to account for the polydispersity in a suspension. James and Ray~\cite{James-SR-2017}
investigated this problem for suspensions in two and three-dimensional
turbulent flows. Interestingly, the authors were able to derive the typical
impact velocities between particle pairs which, in principle, could have
different Stokes numbers. This was conveniently done by assuming a reference
particle with a Stokes number $St_1$ and a second one with $St_2$; in the usual
mono-disperse problem, $St_2 = St_1$. Assuming the smoothness of the underlying
fluid velocity at small inter-particle separations, the impact velocity
$\Delta$ was theoretically calculated (under suitable approximations) for pairs
of droplets with different Stokes numbers and validated against numerical
simulations.

\begin{figure}
\includegraphics[width=1\columnwidth]{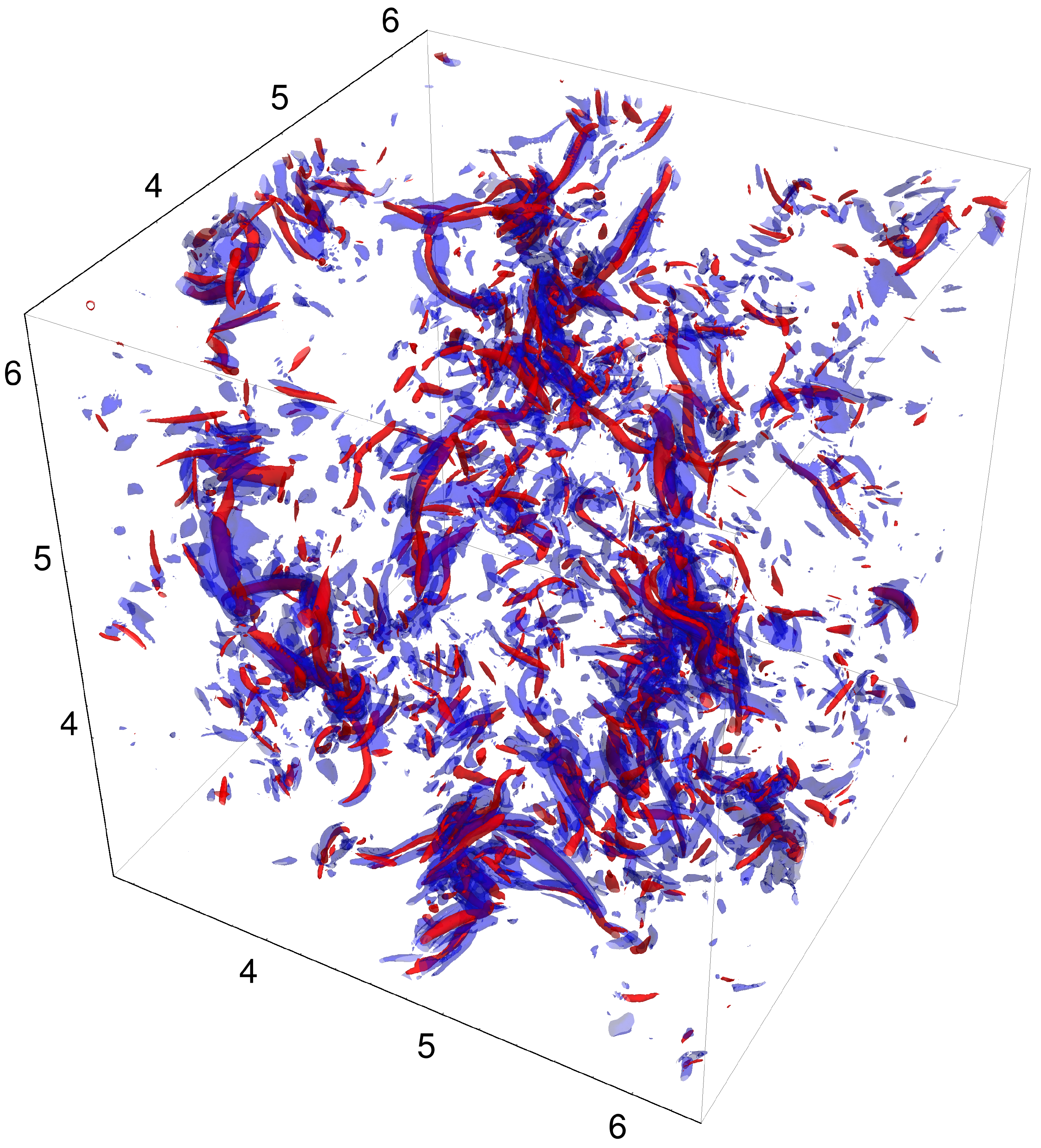}
\caption{Snaphot of intense rotational (red) and straining (blue) regions in a turbulent flow, visualized using the $\mathcal{Q}$-criterion. Intense vortical/rotational regions take the form of tubes or \textit{worms} that are enveloped by strongly straining sheet-like structures, forming vortex-strain wormrolls. Reproduced from~\citep{Picardo_PRF2019}.}
\label{Qstruc}
\end{figure}

As we discussed before, because of the collision-coalescence processes, it is
safe to assume that particle dynamics in a cloud may well be non-stationary. As
a result of this James and Ray have also looked at the collision rates in a
non-stationary phase and found an enhancement of this, when compared to the
collision rates in the statistically stationary phase for all non-zero Stokes
numbers. Interestingly, this ratio peaks to 2 (Fig. 4 in Ref.~\cite{James-SR-2017})
when the Stokes number of the colliding pairs are around 0.2. Although the
results obtained suggests a lack of universality, this observation might be one
possible explanation of possible run-away processes which explains the rapid
growth of droplets from tiny nuclei, seed rain drops. We also refer the reader 
to recent studies in Refs.~\cite{Gustavsson-2011-PRE,Gustavsson-2013-PRE,Meibohm-PRE-2017,Akshay-PRE-2018a,Akshay-PRE-2018b} on the issue of such poly-disperse suspensions and the 
relative velocities of colliding inertial particles in turbulent flows.

\begin{figure*}
\includegraphics[width=1.0\textwidth]{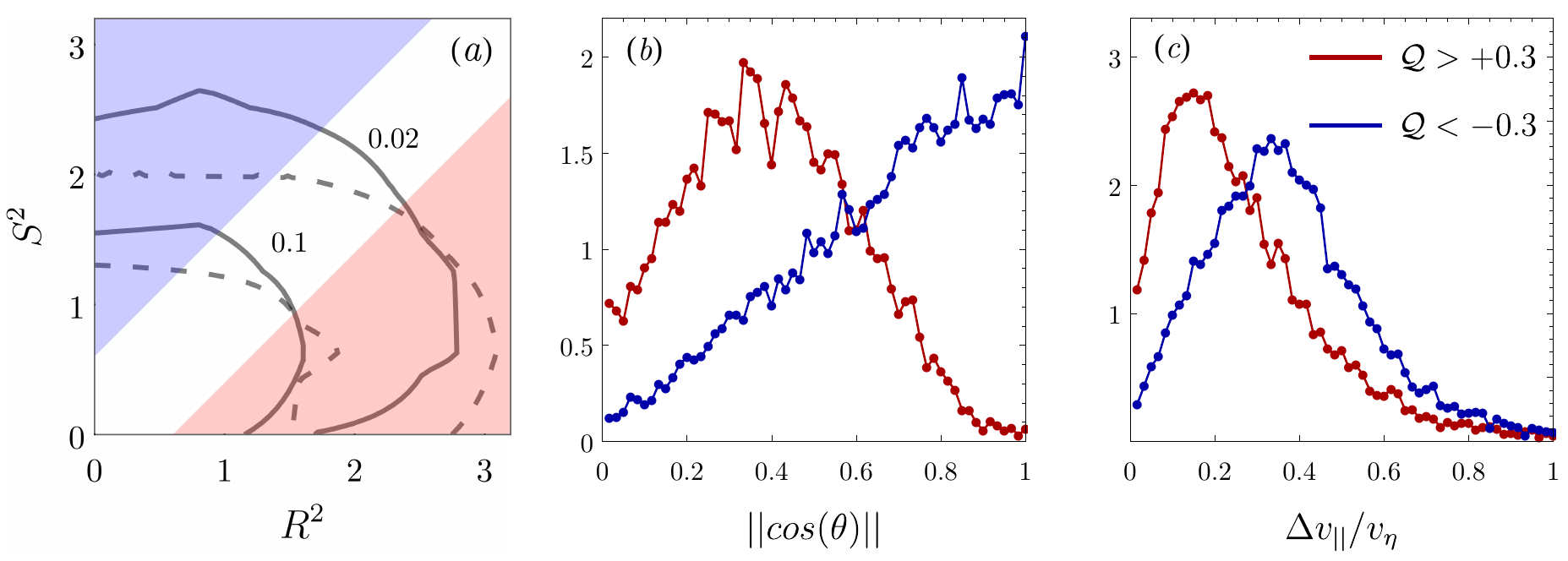}
\caption{Panel (a) presents the 0.1 and 0.03 level contours of the joint probability distribution function of the values of $R^2$ and $S^2$, measured at the positions of inertia-less ($ St = 0$) particles (dashed) and at their collision locations (solid). Panels (b) and (c) present distributions of the collision angles and the collision velocities, respectively, conditioned on whether the collisions occur in rotational ($\mathcal{Q} > +0.3$, red) or straining ($\mathcal{Q} < -0.3$, blue) regions. The contribution of these regions to the joint distribution in panel (a) is shown by the red/blue shading. Taken together, these panels show that collisions in straining regions tend to occur in a head-on or rear-end fashion, which results in a higher collision approach velocity, and thereby a higher collision frequency. Reproduced from~\citep{Picardo_PRF2019}.}
\label{Qang_tracer}
\end{figure*}

All of this inevitably leads us to the question of how the size distribution of droplets 
evolve when we turn on actual coalescences in a suspension advected by a turbulent flow.
A time-honoured theoretical framework for studying such problems is the Smulochowski's 
equation with a stationary coalescence rate or kernel. Starting with an initial infinite bath 
of particles of the same size (and mass), such an approach inevitably leads to a growth 
in the population of droplets of larges sizes as a simple power-law in time. Specifically, 
assuming an initial infinite bath of particles of mass 1, the number of particles with mass $i$ (also, in these units, 
an integer since coalescence imposes mass conservation) grows as 
\begin{equation}
n_i(t) \simeq n_1^i \left({t}/{t_i}\right)^{i-1}.
 \label{eq:evol_mf}
\end{equation}
(The time $t_i$ appearing in the exponent is taken as an 
average time-scales set by the different stationary collision rates 
between all particle pairs which add up to give the $i$-th particle). 

However, when such particles are in a dilute suspension carried by a turbulent
flow, how accurate are these estimates emerging from Smulochowski's equation? 
This question was answered, through a combination of theory and numerical simulations, by 
Bec \textit{et al.} in Ref.~\cite{Bec-PRE-2016}. This work carefully analysed the 
contribution to the coalescence rate coming not from the microphysics of adhesion 
but the fact that particles are in a turbulent flow. This non-trivial contribution
was shown to factor in the anomalous part $\delta_3$ in the scaling of the 
third-order structure function in a turbulence-advected passive-scalar field 
$\langle (\theta(\bm x+\bm r)-\theta(\bm x))^3\rangle \sim |\bm
r|^{1-\delta_3}$
and leads to a population growth of the form 
\begin{equation}
n_i(t) \simeq n_1^i \left({t}/{\tilde{t}_i}\right)^{(1-\frac{3}{2}\delta_3)(i-2)+1}.
	\label{eq:evol_turb}
\end{equation}
Given that the (universal) value of $\delta_3 \approx 0.18$, this form suggests a more rapid  
growth of droplets with masses different from the bath of monomers of mass 1.

The accuracy and correctness of the theoretical calculations were benchmarked
against state of the art direct numerical simulations with an initial
suspension of 1 billion monomers which were then allowed to coalesce and form
droplets of other sizes in the same paper~\cite{Bec-PRE-2016}.  Figure 2 in this
work shows the accuracy of the prediction~\eqref{eq:evol_turb} at early times;
a further confirmation of the importance of anomalous scaling was obtained via
the probability density function of the inter-coalescence times between the
initial monomers (of mass 1) and other droplets of different sizes which were
subsequently formed (Fig. 3 in Ref~\cite{Bec-PRE-2016}). Remarkably, these
results are perhaps the only ones which show how anomalous scaling in
turbulence shows up as a leading order effect in a more applied problem such as
the one of coalescences in turbulent transport.

This work thus established a plausible argument to suggest a rapid growth in
droplets at short times through a complex (Lagrangian) correlated sequence of
events. However the fate of these droplets at long time still remains a large
unexplored issue. 

All of these measurements are of course central in building up
models for collision and coalescences in a warm cloud. However, they do not
help us, in a direct way, to uncover the correlation, if any, between
collisions and the flow structures peculiar to turbulence. Indeed, even
small-scale, homogeneous and isotropic turbulence, which we may expect to
encounter in the core of a cumulus cloud, is rich in structure: It is
perforated by a hierarchy of rotational and straining flow
structures~\citep{she1990,Douady1991,Jimenez1998,Zeff2003,Schumacher2010,davidson2012},
as shown in Fig.~\ref{Qstruc}. These structures are a physical manifestation of
the intermittency of the velocity gradient field, which distinguishes
fully-developed turbulence from a simple random Gaussian
field~\citep{ishihara2007,Ray2009,Tsinober2009}. One may ask, therefore,
whether the collisions between inertial particles or droplets are sensitive to
these flow structures, and thereby to the non-Gaussian nature of turbulence.
This question was addressed recently by Picardo et al.~\citep{Picardo_PRF2019},
who measured the relative values of rotation ($R^2$) and straining ($S^2$) at
the locations of collisions, and compared them to the values sampled by
particle trajectories. They found that collisions among small $St$ particles
are disproportionately frequent in straining regions, much more than what may
be anticipated from preferential concentration alone. In fact, this effect is
not fundamentally tied to inertia, but persists even in the limit of $St \to
0$. for which the particles are homogeneously distributed. (In this limit, the
particles are effectively tracers, but with a small fictional radius that
enables the detection of collisions.)

Figure~\ref{Qang_tracer}(a) presents contours of the joint
probability distribution of the values of $R^2$ and $S^2$, measured both where
inertia-less particles reside (dashed contours) and where they collide (solid
contours). Here, straining (rotational) regions with $\mathcal{Q} < -0.3$ ($>
+0.3$) are shaded in blue (red). Clearly, there there is an oversampling
(undersampling) of straining (vortical) regions by collisions, compared with
the particle trajectories. This discrepancy is a result of the very different
flow geometry in these regions. Colliding particles in straining regions tend
to approach each other in a head-on or rear-end fashion, whereas particles in
rotational regions approach each other in a side-on manner and undergo glancing
collisions. This is shown by the conditioned-PDFs of the collision angle
$\theta$ (the angle between the relative velocity and position vectors of the
two colliding particles), presented in Figure~\ref{Qang_tracer}(b). For a given
magnitude of the underlying fluid velocity, head-on collisions are faster than
side-on collisions, as shown in Figure~\ref{Qang_tracer}(c), because a larger
component of the relative-velocity of the particles is translated into the
collision velocity. For the same number density of particles, this results in a
higher collision frequency in straining regions.

\begin{figure*}
\includegraphics[width=1.0\textwidth]{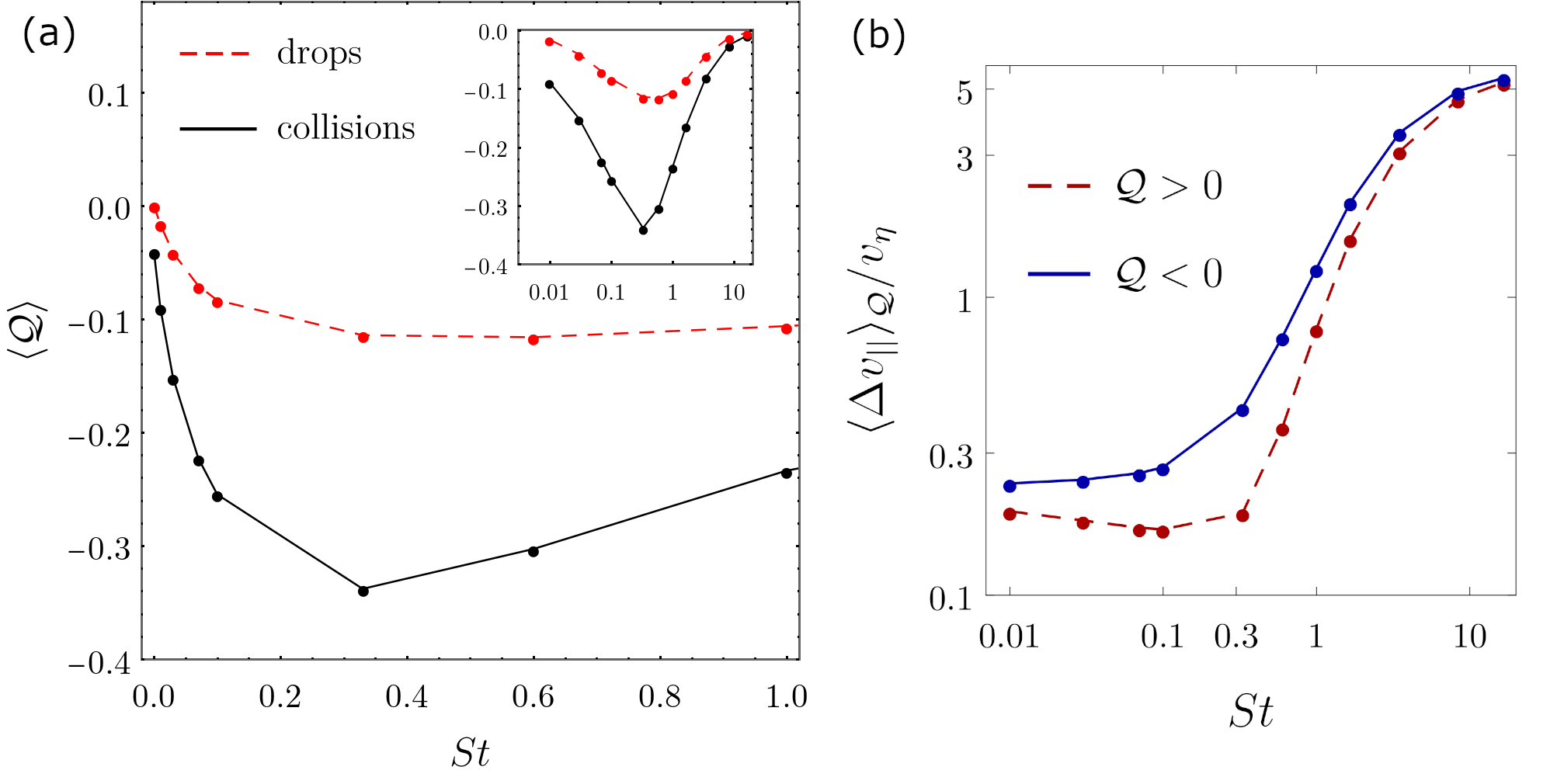}
\caption{Panel (a) presents the average value of $\mathcal Q$ measured where inertial particles or droplets collide (black-solid) and reside (dashed-red), as a function of $St$. The inset presents the same result for a wider range of $St$ using a semi-log scale. Panel (b) presents the average collision velocity, conditioned on whether collisions occur in regions dominated by rotation (red) or strain (blue), as a function of $St$. Adapted from~\citep{Picardo_PRF2019}.}
\label{Qavg_vr}
\end{figure*}

Particle inertia acts to enhance this preference of collisions for straining
regions, upto $St \approx 0.3$. This is seen in Fig.~\ref{Qavg_vr}(a), which
presents the average value of $\mathcal Q = (R^2-S^2)/2$ measured at particle
and collision locations, as a function of $St$. At small $St$, inertia
selective enhances the collision velocity in straining regions, as shown in
Fig.~\ref{Qavg_vr}(b), and therefore increases the frequency of collisions in
straining regions relative to rotational regions. At larger values of $St$,
however, the particles begin to decorrelate from the underlying flow structures
and both particle and collision locations begin to distribute uniformly  (cf.
the inset of Fig.~\ref{Qavg_vr}(a)) and also collide with comparable velocities
in straining and vortical regions (Fig.~\ref{Qavg_vr}(c)). This mismatch
between where inertial (for small $St$) particles reside and collide was
earlier observed by Perrin and Jonker~\citep{Perrin2014}, who also analyzed the
influence of flow structures on collisions using the eigenvalues of the
velocity gradient matrix~\citep{Perrin2016}, which enable a finer
classification of structures than the simple $\mathcal{Q}$-criterion. 

We have seen that collisions between small $St$ particles are
sensitive to the local underlying structure of the flow. As $St$ approaches
unity, however, collisions are affected not just by the structure at the
collision location, but by all the flow structures encountered by the
particles, upto a time of about $\tau_p$ prior to the collision, or upto a
distance of $|v|\tau_p$ around the collision. This raises the possibility of
intense vortical and straining regions conspiring to generate violent, rapid
collisions, due to their peculiar vortex-strain worm-roll geometry (cf.
Fig.~\ref{Qstruc}): Particles in intense vortex tubes will be ejected with
large slip velocities into the enveloping straining sheets, where they have a
high chance of colliding with large relative velocities. Picardo et
al.~\citep{Picardo_PRF2019} found evidence for this scenario, by Lagrangian
backtracking of particles that collided in straining regions: the particles
which collided in the least time after being ejected from a vortex, were indeed
the ones that originated from the strongest vortices, collided in the strongest
straining regions and with the largest collision velocities. 

This effect of vortex-strain worm rolls was found to be
prominent only for $St$ beyond about 0.5. This value may seem too large for
small cloud droplets, and indeed it is when one considers the particle
relaxation time relative to the mean Kolmogorov time-scale of the turbulent
flow. However, at the extremely large $Re$ of in-cloud turbulence, there are
likely to be a few, very intense, intermittent vortices, with a local flow time
scale that is much smaller than the mean Kolmogorov time-scale. This means that
local,\textit{ effective} $St$ of particles in the vicinity of such intense
vortices will be closer to unity, making the vortex ejection and collision
scenario relevant. Even if such intense vortices occupy a very small volume
fraction of the flow, the rapid collisions generated will be able to act as a
seed that initiates the run-away growth of droplets by gravitational driven
collision-coalescence~\citep{Shaw2005}.

\begin{figure}
\includegraphics[width=1\columnwidth]{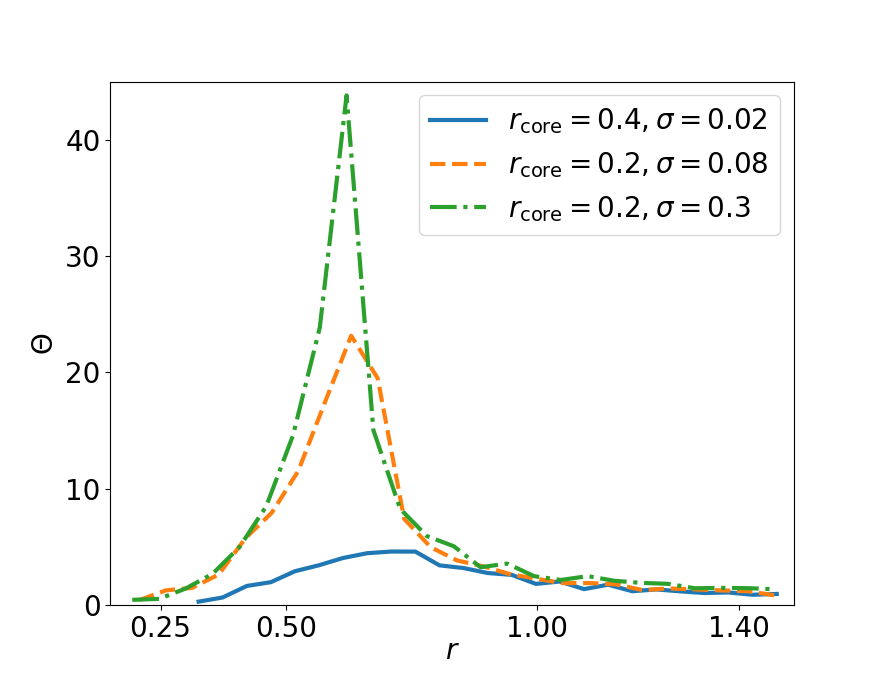}
\caption{Coarse-grained collision density $\Theta$, i.e. the number of collision per unit volume, as a function of the radial distance $r$ from the axis of a tubular Burgers vortex. $r_{core}$ is a measure of the size of the core of the vortex, i.e. of how intensely the vorticity in concentrated about the vortex-axis. $\sigma$ is a measure of the straining flow, that is directed inward along the radial direction and outward along the vortex-axis. This straining is essential for maintaining a concentrated vortex tube in  a viscous fluid, where vorticity continuously diffuses outward~\citep{Davidson}. Here, we see that this straining flow also acts to enhance the collisions around intense vortices. Reproduced from~\citep{Agasthya2019}.}
\label{Burgers}
\end{figure}

Unfortunately, the $Re$ values that can be directly simulated
on a computer are still orders of magnitude smaller than what is expected for a
cloud. Therefore, it is not possible to directly investigate the effect of
high-$Re$ flow structures. However, one can gain a basic understanding of how
such structures may influence the motion and collisions of droplets by using
model vortex flows. Such an analysis was carried out in two-dimensions, using
point and gaussian vortices, by Ravichandran et al.~\citep{Ravichandran2015}
and Deepu et al.~\citep{Deepu2017}, and later extended to a three-dimensional
Burgers vortex~\citep{BURGERS1948171} by Agasthya et al.~\citep{Agasthya2019}.
These studies show that particles near the core of a strong vortex are ejected
more rapidly than particles farther away. This leads to a large increase in the
local particle density around the periphery of the vortex (Fig. 3
of~\citep{Agasthya2019}), as well as large relative velocities between
neighbouring particles (caustics). These factor combine to significantly
enhance collisions in the vicinity of the vortex. Figure~\ref{Burgers} shows
the coarse-grained collision density $\Theta$ (obtained from a large ensemble
of simulations) as a function of the radial distance $r$ from the axis of a
Burgers vortex, which serves as a model for the intense vortex
tubes~\citep{GIBBON1999497,Davidson} observed in three-dimensional turbulence
(cf. Fig.~\ref{Qstruc})~\citep{Douady1991,she1990}. Three cases are presented,
corresponding to a mild, wide vortex ($r_{core} = 0.4$) and an intense thin
vortex ($r_{core} = 0.2$), with mild and strong axial straining ($\sigma =
0.08$ and $0.3$ respectively).  The straining flow along the vortex-axis is
seen to enhance the number of collision produced around an intense vortex, by
drawing particles in towards the core of the vortex.

Taken together, these results show that collisions are definitely sensitive to
the structure of the turbulent flow, and suggest that the intense and
intermittent vortex-strain structures, characteristic of turbulence, may play a
key role in the coalescence driven growth of droplets in clouds. 

Before we conclude this section, let us touch upon briefly the question of how such 
droplets settle under gravity. The mean settling velocity $V_s$, i.e., the component of the particle velocity along the direction of gravity, 
of a particle with a Stokes time $\tau_p$ is simply \eqref{eq:stb}  $V_s = \tau_p g - \langle
u_z({\bf x}_p,t)\rangle$. In the absence of a flow or a uniform sampling of the flow by the particles, 
$\langle u_z(x_p,t)\rangle = 0$ and leading to a predictable settling velocity $V_s = \tau_p g$.
However in the presence of a background turbulent flow, it has been known that
the settling velocity can be enhanced through an oversampling of the regions
where the fluid velocity is downwards. A systematic and quantitative
understanding of this phenomenon was carried out by using extensive numerical
simulations and theory by Bec \textit{et al.} in Ref.~\cite{Bec-PRL-2014}. 

A convenient way to estimate this enhancement of the settling velocity is to measure 
the relative increase $\Delta_V = (V_s-\tau_p g) /
(\tau_p g) = - \langle u_z({\bf x}_p,t)\rangle/ (\tau_p g)$ as a function of the 
Stokes number. In Fig. 2 of Ref~\cite{Bec-PRL-2014}, the authors showed the $\Delta_v$ is 
indeed positive and a non-monotonic function of $St$ with a peak at $St \sim 1$. This enhancement, 
in the limit of small values of $St$ was understood by showing that the correlation 
$\left\langle u_z\nabla_\perp\cdot {\bf v} \right\rangle = \tau_p^2 g \left\langle (\partial_z u_z)^2 \right\rangle >0$. 
This shows that clustering of particles on any horizontal plane (perpendicular to the direction 
of gravity)---and hence  $\nabla_\perp\cdot {\bf v} < 0$---occur at points in the flow where the 
fluid velocity is downwards since for the overall correlation function to be positive $u_z < 0$. Such an 
asymptotics also suggests that for small Stokes numbers $\Delta_v \propto St$. 

The large Stokes asymptotics, dominated by the ballistic motion of particles
resulting in a short-time correlation with the fluid velocity being sampled by
the particle, is more involved. However, the remarkable thing about this
asymptotics is it makes a scaling prediction on $\Delta_v$ in terms of the
Reynolds, Froude and Stokes numbers which are shown, through numerical
simulations (Fig. 2 in Ref.~\cite{Bec-PRL-2014}, to be exceptionally accurate. 

From the perspective of warm clouds, the nature of setting of droplets under
gravity has one further important consequence. Bec \textit{et
al.}~\cite{Bec-PRL-2014} showed that gravitational settling, especially when the
effect of gravity is pronounced, is accompanied by a
quasi-two-dimensionalisation of the particle dynamics (Fig. 3 in
Ref.~\cite{Bec-PRL-2014}).  A consequence of this is the estimation of the
collision rate $\kappa \sim r^\gamma$ which is the average longitudinal
velocity differences between pairs of same-sized particles at a separation $r
\sim 2a \ll \eta$ as they approach each other. It was shown that since  $\gamma
= \xi_1+D_2-1$, where $\xi_1$ is exponent of the order-1 structure function
constructed from particle velocities, the approach rates must be influenced by
the nature of clustering $D_2$ brought about through gravitational settling.
Figure 4 in Ref.~\cite{Bec-PRL-2014} summarises these exponents with their
dependence on both the Stokes and Froude numbers and shows how, under the
influence of gravity, inter-particle approach velocities are diminished,
through a renormalised effective Stokes number, as the effect of gravity begins
to dominate. Indeed, these results suggest that for 30$\mu m$-sized water
droplets, typical in warm clouds, collision rates are almost doubled when we
factor in the interplay of both gravitational and turbulent effects on their
mixing.  These ideas of settling are now being extended to spheroidal particles
in turbulent flows which serve as an effective model for ice crystals in colder
clouds~\cite{Roy2018,anand}.

The findings above are obtained at moderate Reynolds number, while the Reynolds
numbers in a cloud are several orders of magnitude higher. We highlight the
importance of understanding how flow structures and other features change with
increasing Reynolds numbers. Moreover buoyancy effects can be important, as
discussed below.

\section{Microphysics with thermodynamics}

In the previous section we discussed how turbulence affects the dynamics of
droplets, in that it clusters them into portions of the flow, thus encouraging
droplet collisions and merger. We assumed that droplets do not affect the
turbulence. Mechanically speaking, this is a fair assumption through most
stages of droplet growth. This is because water droplets are very small, and
form a very dilute suspension, in that occupy only a millionth of the volume in
a cloud. However, droplets can distort the turbulence that drives them through
the thermodynamics associated with phase change.  Some of the physics behind
these effects was explained in Ravichandran and
Govindarajan~\citep{ravichandran2017b}. 

\subsection{Thermodynamics of phase-change \label{subsec:Thermodynamics-of-phase-change}}

The condensation of water vapour into water and the evaporation of
liquid water into vapour (hereafter just phase-change) are governed
by the Clausius-Clapeyron law, 

\begin{equation}
\frac{dp_{s}}{dT}=\frac{L_{v}p_{s}}{R_{v}T^{2}},\label{eq:clausius}
\end{equation}
where $p_{s}$ is the equilibrium water vapour pressure at the temperature
$T$, $L_{v}$ is the enthalpy of vaporisation, and $R_{v}$ is the
gas constant for water vapour. The Clausius-Clapeyron law can be derived
from the condition that the vapour-liquid system is at equilibrium
at the given temperature (see, e.g. \cite[Chapter 5]{bohren1998}).
This equation can be integrated assuming $L_{v}$ and $R_{v}$ are
constants (this is a reasonable assumption) to give
\begin{equation}
p_{s}=p_{s}^{0}\text{exp}\left(\frac{L_{v}}{R_{v}}\left[\frac{1}{T_{0}}-\frac{1}{T}\right]\right).\label{eq:clausius_integrated}
\end{equation}
Further approximation is possible for small temperature changes $(T_{0}\approx T)$
to give
\begin{equation}
p_{s}=p_{s}^{0}\text{exp}\left(\frac{L_{v}\left(T-T_{0}\right)}{R_{v}T_{0}^{2}}\right).\label{eq:clausius_approx}
\end{equation}
Due to its exponential nature, the amount of water vapour that can
exist in equilibrium is a rapidly changing function of the ambient
temperature; the equilibrium vapour pressure roughly doubles for every
10K increase in temperature.

While the Clausius-Clapeyron law governs the \emph{equilibrium} vapour
pressure, it says nothing about \emph{how} this equilibrium is to
be reached. Chemical reactions or changes of phase that are thermodynamically
favoured may nevertheless not occur because the reactions have high
energies of activation. As a result, pockets of air with higher concentrations
of water vapour than given by equation \ref{eq:clausius_integrated}
are very commonly found in the atmosphere. The system of air and water
vapour is then said to be `supersaturated'. In fact a cloud is often on average supersaturated. So excess water vapour is available, which can then condense. However,  thermodynamically for spontaneous condensation, i.e., condensation without any pre-existing nucleation site, we need about 400\%  supersaturation, whereas such supersaturation is impossible under atmospheric conditions, where it is almost never greater than 5\%. 
We therefore need cloud condensation nuclei on which condensation can occur, and droplets and aerosol particles provide such surfaces. These nuclei are typically small particles of salt or
dust and a background concentration of these nuclei of about $100-1000\text{ cc}^{-1}$
exists in the atmosphere. This number concentration is a function
of how polluted the air is, typically being larger over the continents
than over the ocean. This number concentration, then, also decides
how supersaturated the air can be. Supersaturations for polluted air
are typically $1\%$ or lower, while higher supersaturations are seen
in marine clouds (see, e.g. \cite{pruppacher_klett}, chapter 2). Nuclei
smaller than a critical size (called the Kohler radius) reach an equilibrium
radius and do not grow beyond this (due to the fact that the saturation
vapour pressure is a function of the radius). Nuclei that are larger
than the Kohler radius grow to become water droplets in clouds. These
water droplets continue to grow by absorbing the water vapour in the
atmosphere. The resulting release of the latent heat of vaporisation
drives the large scale dynamics of clouds, as we sketch below.

A relation describing the rate at which the water droplets grow and
consume water vapour can be derived assuming the water droplets are
small enough for ventilation effects to be negligible (see \cite{bohren1998}
chapter 7, \cite{pruppacher_klett}, chapter 13). This gives
\begin{align*}
a\frac{da}{dt} & =\frac{s-1}{C\rho_{w}},
\end{align*}
where $C=\mathcal{O}\left(10^{7}\right)ms/kg$ is a thermodynamic
constant which is a function of the ambient temperature. The rate
of growth of a droplet is inversely proportional to its radius. As
we have seen in section \ref{subsec:particles_caustics_collisions},
this is one of the factors that makes explaining rain-formation challenging.
If this relation is applied to a system of $n$ droplets per unit
volume, ignoring interactions, the rate at which the water vapour
in the system is consumed is
\begin{align}
\frac{d\rho_{v}}{dt} & =-\frac{\rho_{v}/\rho_{s}-1}{\tau_{s}},\label{eq:vapour_phase}
\end{align}
where $\tau_{s}$ is a time-scale and $\rho_{s}=p_{s}/(R_{v}T)$ is
the saturation vapour density. This condensation of vapour results
in the heating of the flow at a rate 
\[
\frac{dT}{dt}=\frac{L_{v}}{C_{p}}\left(\frac{\rho_{v}/\rho_{s}-1}{\tau_{s}}\right).
\]
The latent heat of vaporisation, $L_{v}\approx2.5\times10^{6}J/kg/K$
is a large value. As a result of this, despite the small amounts by
weight of water vapour and liquid found in clouds (typical values
are $\mathcal{O}\left(1-10\right)g/kg$), the amounts of heat released
are enormous and can be $O(MW/m^{3}$ (see, e.g. the discussion
in \cite{RN2011PNAS} on typical heating rates in clouds). The heating
thus provided increases the temperature, and the resulting buoyancy
drives the upward flow of the cloud. Differential heating of regions
of the flow and the resulting buoyancy differences drive the turbulence
in the flow. We look next at how particle inertia, phase
change and buoyancy interact in clouds.

\subsection{Interactions of particle inertia, thermodynamics, and buoyancy-driven
flow \label{subsec:Interactions}}

We refer again to the box-diagram in fig. \ref{fig:interactions}
showing the different interacting phenomena in clouds. All clouds
are composed of water vapour, water droplets, and aerosol particles
suspended in turbulent flow. Particle inertia, phase-change, and buoyancy-driven
turbulent flow are all active in clouds. However, depending on the
type of cloud and the range of parameters, simplifying approximations
may be made which ignore one or more of these effects. We discuss
some relevant examples below which illustrate this.

\subsubsection{Cumulus clouds: phase-change+buoyancy+turbulence \label{subsec:Cumulus}}

Cumulus clouds are tall, heap-like clouds found in the atmosphere,
and are crucial to the maintenance of heat and mass balance in the
atmosphere. Their importance in the dynamics of the atmosphere has
long been recognised, with competing attempts to model them as
various free-shear flows like jets, plumes, or thermals (\cite{stommel1947, squires1962}).
These clouds, driven by the release of latent heat in the flow, differ significantly in their dynamics from jets
and plumes without such latent heating, and have been an object of
study for 60 years. A fuller review of these efforts may be found in \cite{derooy2013}.
The parameter of interest in the study of these clouds is the entrainment rate---the 
rate at which the cloud drags in ambient (dry) air from its environment. Entrainment 
dilutes the cloud, leading to its ultimate demise. Entrainment in free-shear flows is
still a topic of active research, and the addition of volumetric (i.e. not at the source on the ground) heating 
complicates the picture. 

\begin{figure}
\includegraphics[width=1\columnwidth]{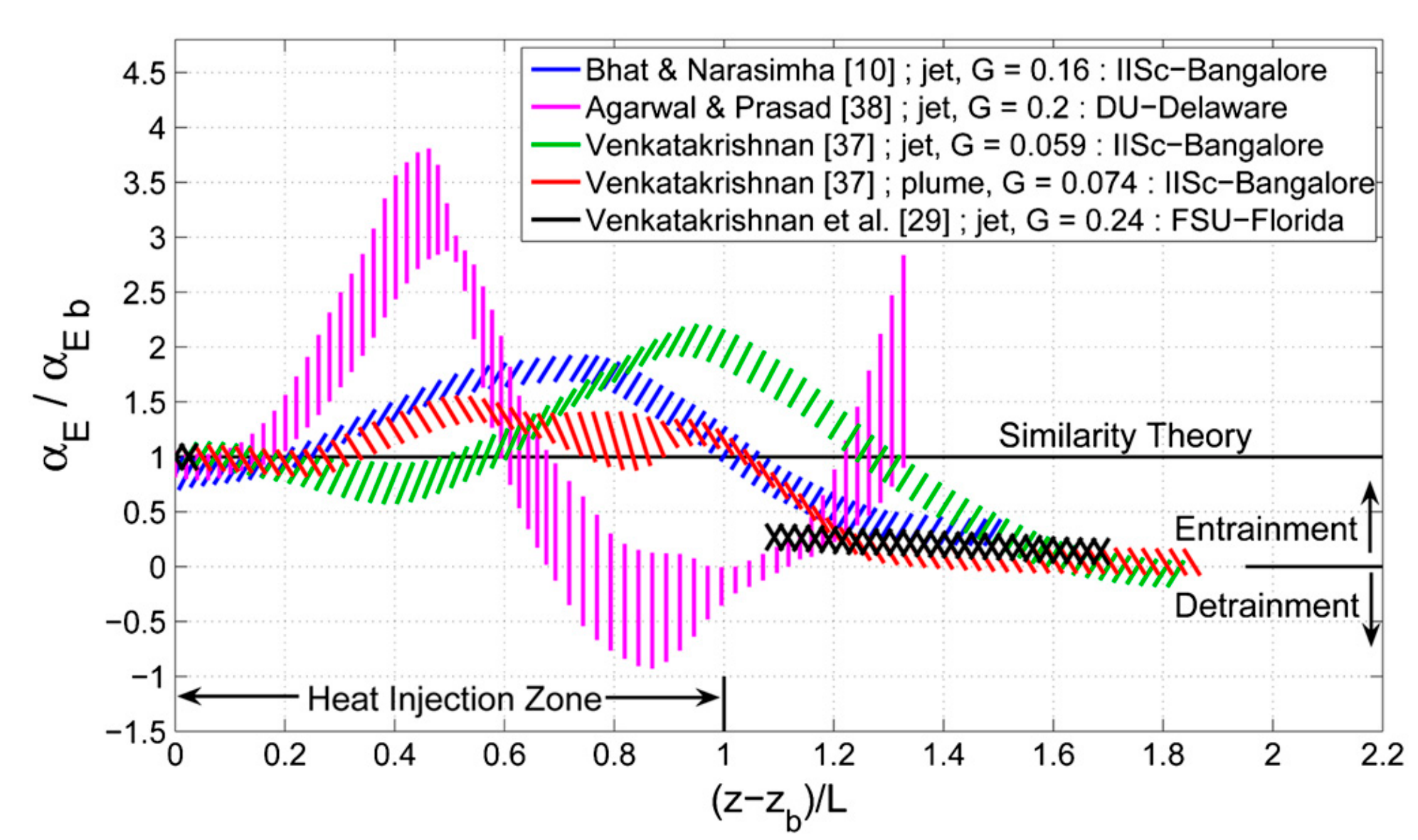}
\caption{The entrainment coefficient in heated plumes first increases and then decreases to zero (i.e.
the plume stops entraining). Reproduced from \cite{RN2011PNAS}. }
\label{entrainment_RN}
\end{figure}

Progress has been made through laboratory experiments on cumulus clouds, showing the role of the latent heat release,
reported by \cite{RN2011PNAS}, building on work in \cite{Bhat1996, venkatakrishnan1998, venkatakrishnan1999}. 
The addition of latent heat to the flow seems to not only accelerate the flow but to shut down the entrainment
of ambient air into the bulk of the flow. This shutdown of entrainment (shown in figure \ref{entrainment_RN}) 
is argued to be because the heating disrupts the coherent structures in the shear layer of the flow (\cite{Bhat1996,RN2011PNAS}). This
then leads to the cloud remaining undiluted for longer and reaching
higher altitudes than if the flow were a pure plume or jet. 

As we have argued, the heating in the cloud arises out of the condensation
of water vapour onto liquid water droplets in the cloud, and thus
a full description of the dynamics would include all the phenomena
listed in figure \ref{fig:interactions}. However, especially for
growing cumulus clouds which have not reached the precipitating state,
the droplet size may be assumed to be small enough that the particle
inertia is negligible. This is a reasonable assumption since the size
distribution in non-precipitating cumulus clouds peaks at $10\mu$m.
This being the case, the droplet inertia is small and the droplets
follow the velocity of the fluid. This in turn allows the droplets
to be coarse-grained into a liquid-water field, a step that saves
significant computational effort. Despite the challenges involved,
progress has been made because of the simplifying assumptions discussed.
Large eddy simulations of cumulus clouds showing this behaviour have
been reported by \cite{Romps2010undiluted}. Laboratory experiments
of cumulus clouds have been reported by \cite{RN2011PNAS}. Direct
numerical simulations for achievable Reynolds numbers are reported
in \cite{ravichandran_narasimha2019}.\\

\subsubsection{Mammatus clouds: phase-change + particle settling + buoyancy\label{subsec:Mammatus}}

A type of cloud that is perhaps not as consequential as the cumulus
clouds, but no less fascinating, is the mammatus cloud. Typically
found underneath cumulonimbus anvil outflows, and therefore acting
as harbingers of inclement weather. The reasons for the formation
of these clouds is a matter of ongoing research, and a comprehensive
review of the various proposed mechanisms may be found in \cite{kanak2006mysteries},
with follow-up studies in \cite{kanak2006idealized, kanak2008numerical}. A promising
explanation involves the combination of the settling of water droplets
out of the cumulonimbus anvils, their subsequent evaporation forming
a layer of air below the anvil that is denser than the ambient air,
and the eventual instability due to this density inversion.  Unlike in
section \ref{subsec:Cumulus}, particle settling cannot be neglected.
A study by \cite{Mammatus2019} finds that the size of the mammatus lobes
is proportional to the settling velocity (which increases for small
droplet sizes like the square of the droplet size) and to the time-scale
for phase-change (which is proportional to the inverse of the mean
droplet size and the number concentration).

\subsubsection{Droplets in clouds, redux: particle inertia+turbulence+phase-change
\label{subsec:Droplets++}}

The phenomenon of warm-rain initiation and the droplet-growth bottleneck
that has been a long-standing unsolved problem in cloud physics, as discussed
in section \ref{subsec:particles_caustics_collisions}. In
studies focussing on the fluid mechanics of droplet collisions-coalescence,
the effects of phase-change and thermodynamics are typically neglected.
In the regime of interest---when droplets have grown to large enough
sizes that their growth rates are small---this assumption is justifiable.
For most of the lifetime of a cloud droplet, however, the thermodynamics
of condensation cannot be neglected. The interactions of phase-change
and particle inertia are thus relevant in the dynamics of clouds:
broad droplet size distributions are important in the rapid growth
of falling droplets through coalescence with smaller droplets. \\

A first step in understanding the interactions of particle inertia
and thermodynamics was taken by \cite{shaw1998} who study how droplets
interact with vortices. Clouds, being turbulent flows, are a tangle
of strong vortices. Vortices, then, are suitable models for idealised
studies. As we have seen in section \ref{subsec:particles_caustics_collisions},
vortices expel inertial particles. When these inertial particles are
also nuclei for condensation, the cores of vortices are voided of
nuclei for condensation and therefore have higher vapour concentrations
than the outside. \cite{shaw1998} argue that this should have two
consequences: first, that any droplets that remain trapped in the
vortical region will end up experiencing larger-than-average supersaturations,
and thus be able to grow to sizes not predicted by only considering
cloud-average values of supersaturation; second, that the supersaturations
produced in the cores of the vortices should lead to the nucleation
of condensation nuclei well above cloud-base, which allows a broadening
of the droplet size distributions (on the lower side). The use by
\cite{shaw1998} of this mechanism in explaining warm-rain initiation
has been questioned in the literature (see \cite{grabowski1999comments, vaillancourt2001, vaillancourt2002}),
but the central message that particle inertia and thermodynamics
interact in complex ways remains relevant, in our view. \\

Clouds are, as we have seen, turbulent flows where the turbulence
is driven by the energy provided by condensing water vapour. While
large velocities can be generated by large values of buoyancy, turbulence
itself, as we have also argued, is generated by \emph{spatial inhomogeneities}
in the heating. \cite{ravichandran2017b}propose a model of how this
can be achieved starting from initially homogeneous conditions, thereby
providing a route by which turbulence can sustain itself by `feeding'
on the latent heat of vaporisation. The mechanism builds on the aforementioned
effect of inertial particles being centrifuged out of vortical regions.
This leaves the vortices more supersaturated (as in \cite{shaw1998}, but 
keeping track of temperature changes), but also colder than their
surroundings which have been heated by the condensation of water vapour.
The resulting density inhomogeneities lead to baroclinic torques which
generate vorticity and thus turbulence in the flow. We mention in
passing that the buoyant vortices that result have interesting dynamics
of their own (see \cite{ravichandran2017a}).

\begin{figure}
\includegraphics[width=1\columnwidth]{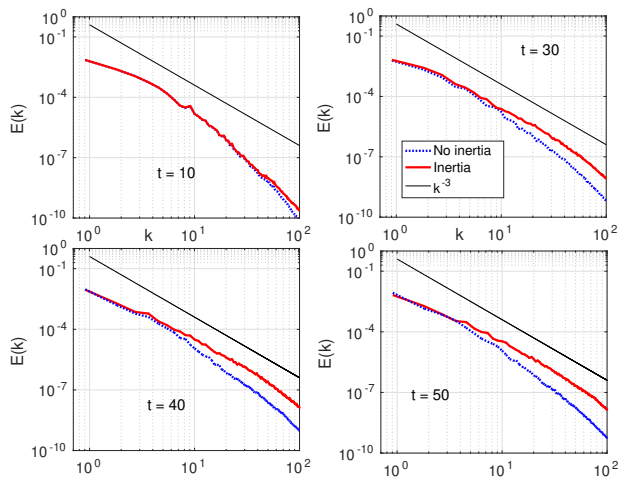}
\caption{The energy density $E(k)$ vs the wavenumber $k$ in simulations
with and without accounting for the effects of particle inertia.
The addition of particle inertia effects provides a route for the
transfer of energy to smaller scales starting from homogeneous conditions. 
Similar to a figure in \cite{ravichandran2017b} }
\label{energy_inertial_vs_tracer}
\end{figure}

Growing clouds are also a class of free-shear flows. The edges of clouds, where
the shear layers separate saturated regions from unsaturated regions,
are thus regions of large inhomogeneities in vapour and droplet concentration.
This results in the generation of strong sustained turbulent flow.
\cite{kumarshaw2012,kumarshaw2013,kumarshaw2014} study this dynamics in
an idealised setup where the temperature is held constant, but water
droplets are allowed to grow or shrink in response to local conditions.
The authors argue that, as a consequence of the high flow Reynolds
numbers in a cloud, the mixing of dry ambient air with the saturated
cloudy air will be highly inhomogeneous: i.e. that parcels of saturated
air and parcels of subsaturated air will often be found close to each
other. As a result, particles in the shear layer can have complicated
growth histories; the size distribution of such droplets will also
be very broad as a result. They quantify this inhomogeneity in terms
of a Damkoehler number which is a ratio of the time-scale of phase-change
to a characteristic timescale of the flow. For large Damkoehler numbers,
which would be expected in the high-Reynolds number flows in the shear
layers of clouds, they show that using a large Damkoehler number $Da \gg 1$ 
results in highly skewed tails with a large
number of droplets that begin at the edge of the shear layer experiencing
extreme evaporation and thus shrinking significantly, while droplets
in the saturated core of the flow grow slightly. In the later stages
of a cloud's evolution, as large droplets formed in the cloud settle
out of the cloud, they are likely to encounter these smaller droplets.
The efficiency of collisions being significantly greater if the droplet
sizes are different, the presence of much smaller droplets in the
shear-layers could play a crucial role in rain formation. \\

More recently, the entire process of rain formation from monodisperse
cloud droplets has been studied by \cite{gotohsaito2016,gotohsaito2018}.
The authors develop a moving box model where they account for the
upward velocity of a cloud by changing the mean background properties
seen by the box. This allows them to follow a subvolume of the cloud
as it ascends upwards. The cloud droplets initially grow by condensation
of vapour. These cloud droplets eventually lead to a small number
of larger droplets by collisions with each other. This is a weak effect,
since collisions of similarly-sized particles are rare. The larger
droplets that eventually form start to settle at significant velocities,
and grow rapidly because of collisions-coalescence with the smaller
droplets. The authors report that the intensity of the turbulence
(which is artificially chosen instead of plays a key role in the collisions-coalescence
process, in line with expectations from work on the problem over the
last two decades.

\section*{Conclusion and Future Directions}

The fluid dynamics of clouds is only part--if the most complex part--of climate science. 
We have argued here that understanding the dynamics is important in the face of the 
uncertainties of climate change. We hope to have convinced readers of the immense challenges and opportunities
the field presents, both as an exercise in scientific curiosity and because of the 
far-reaching implications. Both these facets arise from the range of length- and time-scales
present in the dynamics. The approach we favour---of understanding the parts in service
of understanding the whole---has led to several semi-independent programmes of
research which have to be eventually be united. Under present computational limitations
the goal of these studies is to be able to parameterise the dynamics and improve global climate
models.

Some areas of active research where the questions are reasonably well-defined and can be addressed computationally are
a) A fuller understanding of entrainment in free-shear flows in general, and cloud-flows in particular. Open problems include the magnitude of the entrainment and the details of the process by which volumetric heating alters the entrainment; by extension, entrainment in merging plumes/cumulus clouds as a way of understanding the dynamics of convective aggregation; and the incorporation of the resultant better models for entrainment into global climate simulations: perhaps by extending super-parameterisation method (see \cite{randall2003}).
b) Droplet-resolved simulations for collision-coalescence, in order to overcome limitations of the geometric collisions approach; to include effects of droplet splintering, droplet interactions, and flow-droplet coupling; and to include effects of differences in liquid/gas properties like density, temperature, conductivity in the phase-change process.

\section*{Acknowledgements}

SR is supported under Swedish Research Council grant no. 638-2013-9243.
JRP acknowledges funding from the IITB-IRCC seed grant.
SSR acknowledges DST (India) project MTR/2019/001553 for support.
RG and SSR  acknowledge support of the DAE, Govt.~of~India, under project no.~12-R\&D-TFR-5.10-1100.


\end{document}